# Watt-level all polarization-maintaining femtosecond fiber laser source at 1100 nm for multicolor two-photon fluorescence excitation of fluorescent proteins

JUNPENG WEN,[1,2] CHRISTIAN PILGER,[3] WENLONG WANG,[1,2] RAGHU ERAPANEEDI,[4] HAO XIU,[1,2] YIHENG FAN,[1,2] XU HU,[1,2] THOMAS HUSER,[3] FRIEDEMANN KIEFER,[4,5] XIAOMING WEI,[1,2,6,7,9] AND ZHONGMIN YANG[1,2,6,7,8,10]

[1]*School of Physics and Optoelectronics, South China University of Technology, Guangzhou 510640, China*
[2]*State Key Laboratory of Luminescent Materials and Devices and Institute of Optical Communication Materials, South China University of Technology, Guangzhou 510640, China*
[3]*Biomolecular Photonics Research Group, Faculty of Physics, Bielefeld University, D-33615 Bielefeld, Germany*
[4]*European Institute for Molecular Imaging, University of Münster, Röntgenstraße 16, D-48149, Münster, Germany*
[5]*Max Planck Institute for Molecular Biomedicine, Röntgenstraße 20, D-48149 Münster, Germany*
[6]*Guangdong Engineering Technology Research and Development Center of Special Optical Fiber Materials and Devices, South China University of Technology, Guangzhou 510640, China*
[7]*Guangdong Provincial Key Laboratory of Fiber Laser Materials and Applied Techniques, South China University of Technology, Guangzhou 510640, China*
[8]*Research Institute of Future Technology, South China Normal University, Guangzhou, Guangdong 510006, China*
[9]*xmwei@scut.edu.cn*
[10]*yangzm@scut.edu.cn*

**Abstract:** We demonstrate a compact watt-level all polarization-maintaining (PM) femtosecond fiber laser source at 1100 nm. The fiber laser source is seeded by an all PM fiber mode-locked laser employing a nonlinear amplifying loop mirror. The seed laser can generate stable pulses at a fundamental repetition rate of 40.71 MHz with a signal-to-noise rate of >100 dB and an integrated relative intensity noise of only ~0.061%. After two-stage external amplification and pulse compression, an output power of ~1.47 W (corresponding to a pulse energy of ~36.1 nJ) and a pulse duration of ~251 fs are obtained. The 1100 nm femtosecond fiber laser is then employed as the excitation light source for multicolor multi-photon fluorescence microscopy of Chinese hamster ovary (CHO) cells stably expressing red fluorescent proteins.

## 1. Introduction

Multiphoton microscopy (MPM) [1] has been recognized as a powerful technique for biological applications due to its high spatial resolution and low optical damage [2,3]. MPM has been applied to various research topics, such as monitoring cell structure and function, investigating the dynamics of cellular networks, mapping the brain activity of animals *in vivo*, and studying fundamental mechanisms of diseases [4–9]. Near-infrared femtosecond lasers are key to MPM because of their deep tissue penetration and weak phototoxicity [10,11]. A variety of fluorescent probes are available for two-photon excitation at near-infrared wavelengths [12–15], leading to an urgent need for high-performance infrared femtosecond laser sources. Short-pulsed laser sources at 1100 nm are particularly attractive, because at this wavelength, several monomeric red fluorescent proteins (FP) can be excited simultaneously [12]. Two-photon calcium imaging with 1100 nm femtosecond lasers can effectively suppress photo-stimulation artifacts [16,17], creating great potential for deep brain imaging [18]. Despite these exciting opportunities, the maximum emission wavelength of commonly used femtosecond laser sources, e.g., Ti: sapphire femtosecond lasers, is generally limited to ~1040 nm. To generate femtosecond laser pulses near 1100 nm, wavelength shifting is widely utilized by exploring nonlinear effects, e.g., self-phase modulation [19], stimulated Raman scattering (SRS) [20], soliton self-frequency shift (SSFS) [21], etc. However, these methods rely on extra free-space components, which, to some extent, increase the complexity and decrease the reliability [22, 23]. Recently, mode-locked fiber lasers operating at 1100 nm have been investigated by employing nonlinear amplifying loop mirrors (NALM) [24,25], wherein the use of spatial coupling in the fiber link might compromise the robustness.

In this letter, we report a compact and robust watt-level femtosecond all polarization-maintaining (PM) mode-locked fiber laser source at 1100 nm. A Yb-doped mode-locked all PM fiber seed laser at 1100 nm is constructed by utilizing NALM, wherein the amplified spontaneous emission (ASE) is specifically suppressed. The mode-locked pulses at a repetition rate of 40.71 MHz have a signal-to-noise ratio (SNR) of >100 dB and an integrated relative intensity noise (RIN) of only ~0.061%. The seed pulses are amplified by two stages of all PM fiber amplifiers, such that the average output power is boosted to ~1.47 W. The amplified pulses are subsequently compressed by a grating pair, yielding a pulse duration of ~251 fs. The femtosecond laser source was then employed as the excitation light source of MPM, for imaging Chinese hamster ovary (CHO) cells stably expressing red FPs.

## 2. Experimental setup

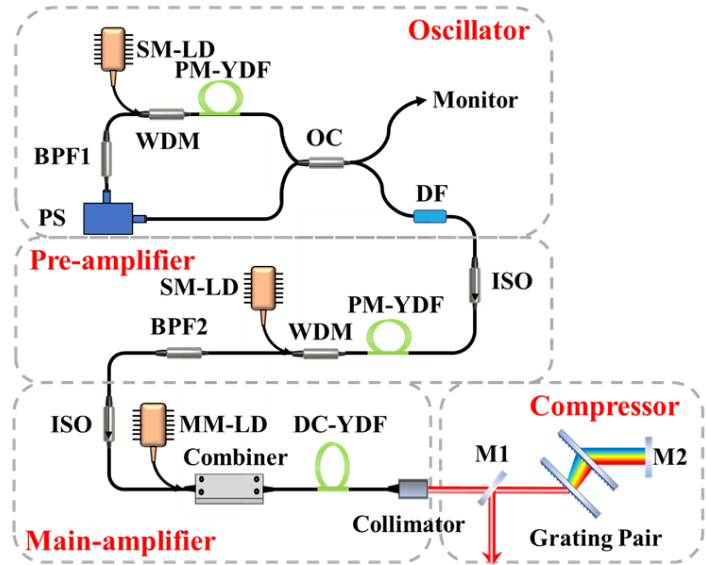

**Fig. 1.** Schematics of the high-power all PM fiber laser source system at 1100 nm. WDM, wavelength-division multiplexer; BPF, band-pass filter; PS, phase shifter; OC, optical coupler; DF, dielectric film; SM-LD, single-mode laser diode; PM-YDF, polarization-maintaining Yb-doped fiber; ISO, isolator; MM-LD, multi-mode laser diode; DC-YDF, double-cladding Yb-doped fiber; M, mirror.

The schematic diagram of the laser source is illustrated in Fig. 1. The seed laser has a typical figure-9 configuration, where the mode-locking is implemented through NALM. The figure-9 seed laser is comprised of a wavelength-division multiplexer (WDM), a band-pass filter (BPF, i.e., BPF1), a $\pi/2$ phase shifter (PS), a 45:55 optical coupler (OC), a fiber-type dielectric film (DF), a 976 nm single-mode laser diode (SM-LD) and a piece of PM Yb-doped fiber (PM-YDF, iXblue, IXF-DF-PM-6-125). The lasing wavelength is confined by the BPF1 centered at 1100 nm which has a bandwidth of 5 nm. The mode-locked seed laser can operate in the self-starting mode-locking state in the all-normal-dispersion regime. The fiber-type DF placed after the OC has a reflectivity of 95% at 1095-1105 nm and a transmittance of 90% at 1010-1075 nm. To enhance the nonlinear phase difference between the clockwise- and counterclockwise-propagating beams inside the fiber loop, a 0.8 m-long PM-YDF, i.e., the gain fiber, is directly spliced to the 55% port of the OC. The gain fiber is core-pumped by a 976 nm SM-LD with a maximum output power of 680 mW through a 976/1100 nm WDM. The pigtails of each device are PM single-mode fibers. The total cavity length is approximately 5 m, corresponding to a fundamental repetition rate of about 40.71 MHz.

The seed laser is protected from back-reflected light by an optical isolator (ISO), and subsequently injected into the amplification system, consisting of two stages of all PM fiber amplifiers. The pre-amplifier has a 3.4 m PM-YDF and is backward core-pumped by a 976 nm SM-LD with a maximum output power of 460 mW through a WDM. Although a relatively long gain fiber is employed to ensure longer-wavelength emission and ASE reabsorption, significant ASE still exists. Therefore, a BPF with a center wavelength of 1100 nm and a bandwidth of 10 nm (BPF2) is placed after the WDM in the pre-amplifier. Another ISO is placed between the pre-amplifier and the main-amplifier to prevent back-reflection. The main-amplifier has a similar configuration as that of the pre-amplifier, except for the pump scheme. The gain fiber of the main-amplifier, i.e. a 3 m-long double-cladding Yb-doped fiber (DC-YDF, Nufern, PM-YDF-10/125-M), is forward cladding-pumped by a multi-mode laser diode (MM-LD) with a maximum output power of 27 W through a signal-pump combiner. The amplified pulses are then launched into free space by a collimator, where a transmission grating pair (LightSmyth, T-1000-1040) with a groove density of 1000 lines/mm serves as the pulse compressor.

### 3. Results and discussion

The output optical spectrum of the laser source is analyzed by an optical spectrum analyzer (YOKOGAWA AQ6370D) with a 0.02nm resolution. The pulse train of the laser source is detected by a high-speed InGaAs photodetector (Newport, 818-BB-51F, 12.5GHz bandwidth) and recorded by a real-time oscilloscope (Teledyne, SDA 820Zi-B). The radiofrequency (RF) spectrum and noise performance are investigated by a frequency signal analyzer (Rohde & Schwarz, FSWP26, 26.5 GHz bandwidth). The power is measured by two photodiode power sensors (Thorlabs, S122C and S146C). The autocorrelation trace is acquired by an autocorrelator (Femtochrome, FR-103XL).

*3.1 Fiber laser source performance*

The basic characteristics of the seed laser are illustrated in Fig. 2. The seed laser operates in the continuous-wave (CW) laser regime at a pump power threshold of 215 mW. By further increasing the pump power to 323 mW, the self-starting mode-locking state is achieved with multiple pulses. By subsequently decreasing the pump power to 230 mW, the seed laser then operates at the stable single-pulse mode-locked state with an average output power of 4.9 mW. Figs. 2(a) and (b) show the optical spectra from the reflection and transmission ports of the NALM, respectively. The optical spectrum from the reflection port has a center wavelength of 1097 nm and a spectral width of 4.7 nm. Obvious ASE noise can be observed in the wavelength range between 1040-1080 nm, and only a SNR of >30 dB is obtained. The pulse train has a temporal spacing of ~24.56 ns [see Fig. 2(c)], corresponding to a fundamental repetition rate of 40.71 MHz [see Fig. 2(d)]. The radio-frequency (RF) spectrum, i.e., Fig. 2(d), is measured at a resolution bandwidth of 10 Hz, wherein an SNR of >100 dB implies good stability of mode-locking. The phase noise (PN) of the mode-locked pulses is also analyzed and shown in Fig. 3(a), and the noise intensity gradually decreases from −100 dBc/Hz to −150 dBc/Hz as the offset frequency increases from 10 Hz to 1 MHz. The integrated timing jitter is ~405 fs. The RIN of the mode-locked pulses is presented in Fig. 3(b), and the integrated RIN from 10 Hz to 1 MHz is only 0.061%.

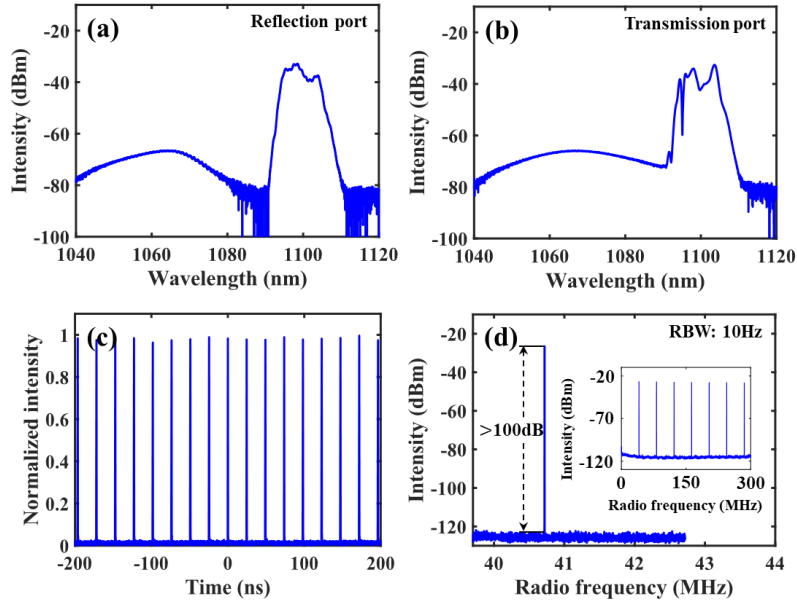

**Fig. 2.** Performance of the seed laser. (a, b) Output spectra from the reflection and transmission ports of the NALM, respectively. The optical spectra are measured with a spectral resolution of 0.02 nm. (c) Mode-locked pulse train recorded by a real-time oscilloscope. (d) Radio-frequency (RF) spectrum of the mode-locked pulses. The inset shows the RF spectrum in a wider span. The RF spectra are measured with a resolution bandwidth (RBW) of 10 Hz.

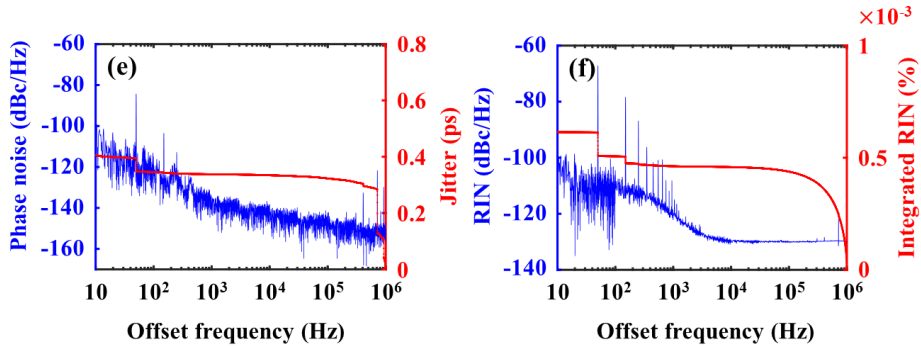

**Fig. 3.** (a) Phase noise (PN) (blue) and integrated timing jitter (red) of the seed laser. (b) Relative intensity noise (RIN) (blue) and integrated RIN (red) of the seed laser.

The seed laser is subsequently fed into the cascaded two stages of the optical amplifiers, i.e., pre-amplifier and main-amplifier. The average output power of the seed pulses is boosted to 100.5 mW by the pre-amplifier at a pump power of 350 mW, corresponding to a net gain of ~13 dB. The pulse duration after the pre-amplifier is measured to be ~5 ps. It is worth noting that the optical spectrum after the pre-amplifier is slightly broadened due to the self-phase modulation effect, and significant ASE noise is generated [see Fig. 4(a)]. Fig. 4(b) shows the optical spectrum after the main-amplifier at a pump power of 2.82 W. The ASE noise at 1040-1080 nm is further increased after the main-amplifier. To remove the ASE noise, a 10 nm BPF (i.e., BPF2) centered at 1100 nm is placed right after the pre-amplifier. After the BPF2, the average output power is decreased to ~79 mW, and an improved SNR of ~60 dB is realized [see Fig. 4(c)]. The optical spectrum after the main-amplifier at a pump power of 4.11 W is shown in Fig. 4(d).

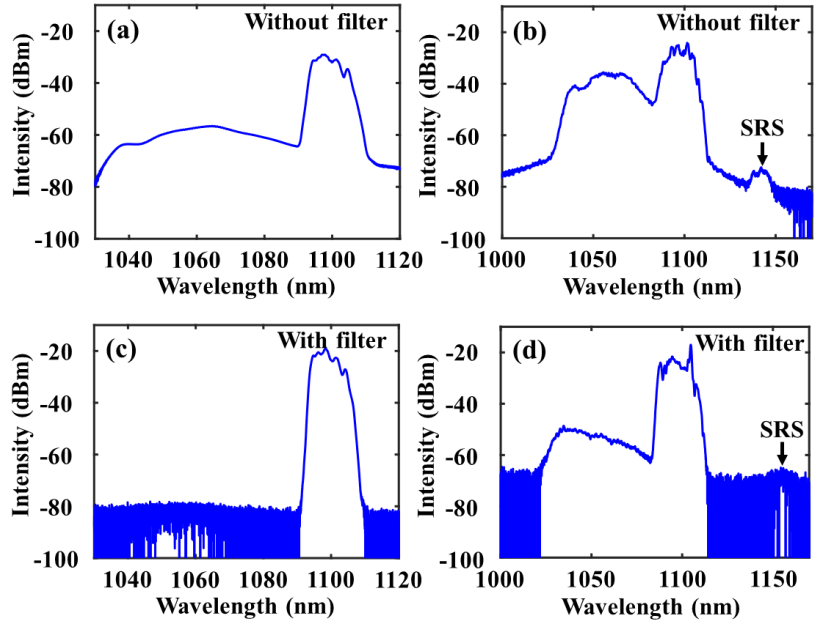

**Fig. 4.** Optical spectra after the pre-amplifier without (a) and with (c) the filter at a pump power of 350 mW. The optical spectra after the main-amplifier without (b) and with (d) the filter at a pump power of 2.82 W and 4.11 W, respectively. SRS, stimulated Raman scattering.

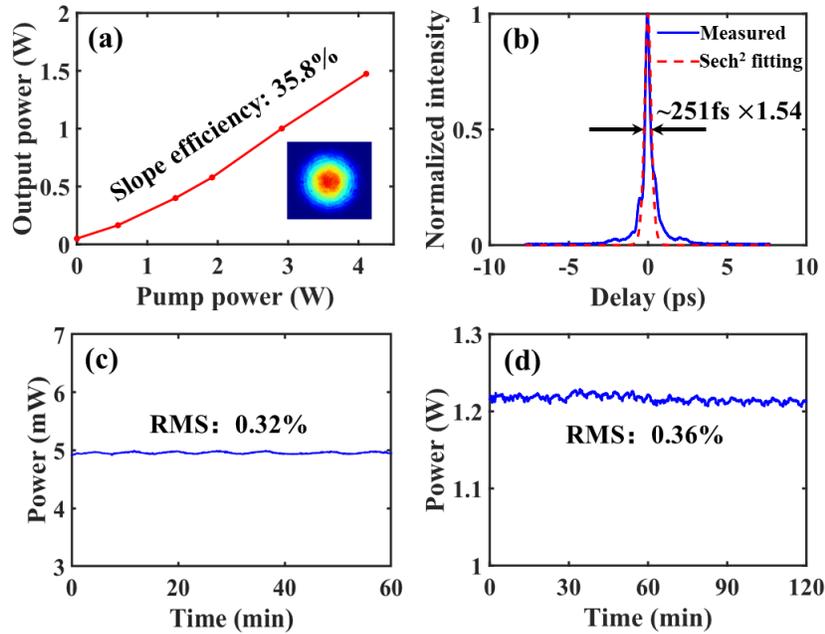

**Fig. 5.** (a) Output power of the main-amplifier as a function of the pump power. Inset: the beam profile. (b) Autocorrelation trace of the amplified pulses after pulse compression, assuming a sech$^2$ pulse shape. The red dashed curve is the fitting. (c, d) Stabilities of the output power of the seed laser over 60 min and the main-amplifier over 120 min, respectively. RMS, root mean square.

After the main-amplifier, the average output power of the amplified pulses reaches ~1.47 W at a pump power of 4.11 W, corresponding to a slope efficiency of 35.8%, as presented in Fig. 5(a). The maximum output power is limited by SRS, as indicated in Fig. 4(d), where the Raman scattering spectral component can be observed at around 1150 nm. It should be pointed out that in the case without the BPF2, the maximum average output power is 1.24 W at a pump power of 2.82 W, corresponding to a slope efficiency of 43.9%. The slope efficiency difference between the two cases can be attributed to ASE. A transmission grating pair is finally utilized to de-chirp the amplified pulses. At a grating-pair separation of ~39 mm, the pulse duration is compressed to about 251 fs, as shown in Fig. 5(b). After the pulse compressor, the maximum output power decreases to 1.03 W. The autocorrelation trace is acquired by an autocorrelator (Femtochrome, FR-103XL). To evaluate the long-term stability of the femtosecond laser source, the output powers of the seed laser and the amplification system are monitored for periods of 60 and 120 minutes, respectively, as shown in Figs. 5(c) and (d). The results suggest that this femtosecond fiber laser source has promising power stability.

*3.2 Multicolor two-photon fluorescence imaging test*

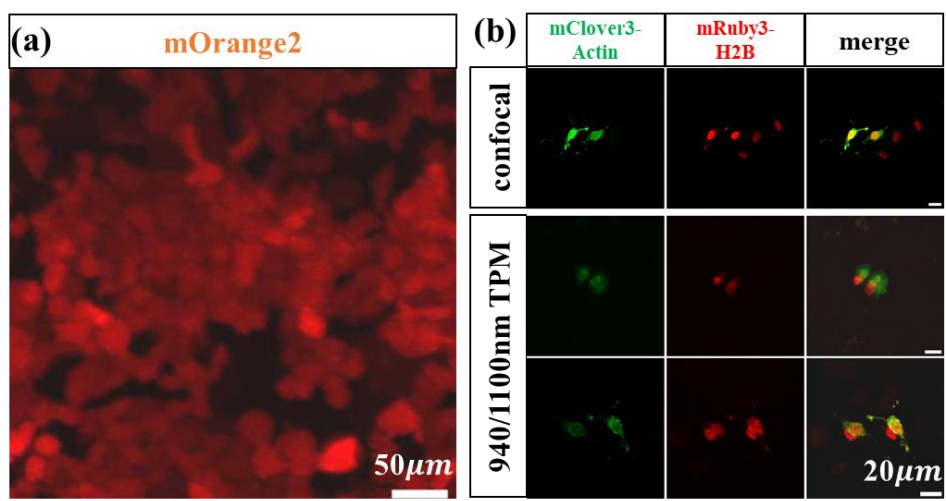

**Fig. 6.** (a) (a) Two-photon microscopy (TPM) fluorescence images of a Chinese hamster ovary (CHO) cell culture stably expressing mOrange2. (b) Two-color TPM fluorescence images of individual examples of CHO cells transiently transfected with mClover3-labelled actin and mRuby3-labelled histone 2B. Top: confocal single-photon microscopy (SPM) images. Middle and bottom: TPM images. Excitation wavelengths: 940 nm (left column, Ti: sapphire laser) and 1100 nm (middle column, fiber laser). s

The 1100 nm fiber laser was then used as the excitation source in combination with a commercial multiphoton laser scanning microscope (TriM Scope Matrix, LaVision Biotec/Miltenyi Biotec) As shown in Fig. 6(a), we demonstrate two-photon excitation of CHO cells stably expressing the red FP mOrange2, which is a representative of the fruit-basket series of fluorescent proteins [26]. Fig. 6(b) shows confocal single-photon microscopy (SPM) images and two-photon microscopy (TPM) images of CHO cells transiently transfected with mClover3-actin (RRID:Addgene_74259) and mRuby3- histone 2B (RRID:Addgene_74258) fusion proteins [27]. The mClover3 and mRuby3 fusion proteins are excited by femtosecond lasers at 940 nm and 1100 nm, respectively, such that two-color TPM fluorescence images can be obtained [Fig. 6(b)], convincingly demonstrating the power of the excitation wavelength range of 1100 nm. Despite efficient excitation of mClover3 at 940 nm, mRuby3 remains dark, while mClover3 is dark at 1100 nm and mRuby3 is efficiently excited. A wide spectrum of green and red FPs including GFP and its derivatives as well as the above-mentioned dsRed and

Entacmaea-derived FP can be efficiently visualized with a two-wavelength femtosecond laser system.

## 4. Conclusions

In conclusion, we have demonstrated a compact all PM fiber femtosecond laser source at 1100 nm. The femtosecond fiber laser source was seeded by a stable mode-locked all PM fiber laser with a 405 fs timing jitter and a 0.061% integrated RIN. After optical amplification and pulse compression, it enables a pulse duration of 251 fs and a pulse energy of 36.1 nJ, with an average output power of 1.47 W. We performed two-photon fluorescence excitation imaging in CHO cells based on this new femtosecond fiber laser source at 1100 nm. Our results suggest that this compact and robust femtosecond fiber laser serves as an efficient source for MPM applications at 1100 nm

**Funding.** This work was partially supported by National Natural Science Foundation of China (62375087, 12374304, U1609219 and 62235014), NSFC Development of National Major Scientific Research Instrument (61927816), Mobility Programme of the Sino-German (M-0296), Introduced Innovative Team Project of Guangdong Pearl River Talents Program (2021ZT09Z109), Natural Science Foundation of Guangdong Province (2021B1515020074), Science and Technology Project of Guangdong (2020B1212060002), and Deutsche Forschungsgemeinschaft (DFG, CRC 1450 – 431460824).

**Acknowledgments.** We gratefully acknowledge support and the gift of transiently transfected CHO cells by Nicole Frost and Nadine Bauer. FK and TH gratefully acknowledge DFG grants INST 211/900-1 FUGG and INST 215/614-1 FUGG for the purchase of their multiphoton microscopes.

**Disclosures.** The authors declare no conflicts of interest.

**Data availability.** Data underlying the results presented in this paper are not publicly available at this time but may be obtained from the authors upon reasonable request.